\documentclass[12pt]{iopart}

\usepackage{epsfig} 
\begin{document}

\title[Energy evolution in time-dependent harmonic oscillator with the arbitrary \ldots]{Energy evolution in 
time-dependent harmonic oscillator with arbitrary external forcing}

\author{Andrei V. Kuzmin$^{1,2}$ and Marko Robnik$^1$ }

\address{$^1$CAMTP --- Center for Applied Mathematics and Theoretical Physics, University of Maribor, Krekova 2, SI-2000 Maribor, Slovenia}

\address{$^2$Joint Institute for Power and Nuclear Research, National Academy of Sciences of Belarus,  Krasina str. 99, 220109 Minsk,
Belarus}

\ead{avkuzmin@sosny.bas-net.by and robnik@uni-mb.si }

\begin{abstract}
The classical Hamiltonian system of time-dependent harmonic oscillator driven by the arbitrary external time-dependent force is 
considered. Exact analytical solution of the corresponding equations of motion is constructed in the framework of the technique 
(Robnik M, Romanovski V G, J. Phys. A: Math. Gen. {\bf 33} (2000) 5093) based on WKB approach. Energy evolution for the ensemble
of uniformly distributed w.r.t. the canonical angle initial conditions on the initial invariant torus is studied. Exact
expressions for the energy moments of arbitrary order taken at arbitrary time moment are analytically derived. Corresponding
characteristic function is analytically constructed in the form of infinite series and numerically evaluated for certain values 
of the system parameters. Energy distribution function is numerically obtained in some particular cases. In the limit of small 
initial ensemble's energy the relevant formula for the energy distribution function is analytically derived.


\end{abstract}

\submitto{\JPA}
\maketitle

\section{Introduction}

In time-dependent Hamiltonian systems the energy is generally not conserved during the time evolution. However, at any given moment of time it has a precise meaning. Namely, it is the value of the Hamilton function calculated for the particular solution of the equations of motion at that moment of time. Thus the system's energy depends both on initial conditions and on time (in the case of time-independent Hamiltonian systems it is specified by the initial conditions only and is conserved). In this paper we restrict ourselves to the consideration of the one-dimensional time-dependent harmonic oscillator subjected to an external forcing. This model arises in a variety of physical problems starting from parametric resonance~\cite{ParamRes} and ending with the geometrical ion trap quantum computations~\cite{GeomTRIO,PRA3}. The one-dimensional harmonic oscillator was considered in a number of works mainly in the context of preservation of adiabatic invariants. After the paper by Einstein~\cite{Einstein}, Kulsrud was the first who showed, using WKB-type method, that for a finite time span $T$ the action variable $I$ is preserved to all orders for the harmonic oscillator, if all derivatives of the oscillator frequency $\omega (t)$ vanish at the beginning and at the end of the time interval, whilst if there is a discontinuity in one of the derivatives he estimated the corresponding change of the action $\Delta I$~\cite{Kulsrud}. The same result was independently obtained for a charged particle in a slowly varying magnetic field for infinite time domain in~\cite{Schluter}. Explicit general expressions were recently derived in~\cite{Robnik2006}. Kruskal, as reported in~\cite{Gardner}, and Lenard~\cite{Lenard} studied more general systems, whilst Gardner used the classical Hamiltonian perturbation theory~\cite{Gardner}. Courant and Snyder have studied the stability of the synchrotron and analysed adiabatic invariant employing the transition matrix~\cite{Courant}. The interest then shifted to the infinite time domain. Littlewood showed for the time-dependent harmonic oscillator that if $\omega (t)$ is an analytic function, then the action $I$ is preserved to all orders of the {\it adiabatic parameter} $\epsilon = 1/T$. Kruskal developed the asymptotic theory of Hamiltonian and other systems with all solutions  nearly periodic~\cite{Kruskal}. Lewis, using Kruskal's method, discovered a connection between the action variable $I$ of the one-dimensional harmonic oscillator and another nonlinear differential equation~\cite{Lewis}. Symon used Lewis'es result to calculate the (canonical) ensemble average of $I$ and its variance~\cite{Symon}. Finally, Knorr and Pfirsch proved that $\Delta I \sim \exp{\left(- c / \epsilon\right)}$~\cite{Knorr}. Meyer relaxed some conditions and calculated the constant $c$~\cite{Meyer1,Meyer2}. Exponential preservation of the adiabatic invariant $I$ for an analytic $\omega (t)$ on $(-\infty , + \infty)$ with constant limits at $t \to \pm \infty$ is thus well established~\cite{Landau}. Energy evolution in a one-dimensional time-dependent harmonic oscillator without forcing was studied by Robnik and Romanovski in~\cite{Robnik2006,RobnikRomanovski}. The exact energy distribution function in this case was also derived~\cite{LastLetter}.

In this paper we consider one-dimensional classical time-dependent harmonic oscillator driven by an arbitrary time-dependent force. The plan is as follows. In section~\ref{Eq} exact analytical solution of the corresponding equations of motion is constructed in the framework of the previously developed technique based on WKB approach~\cite{WKBMod}. Energy evolution for the ensemble of uniformly distributed (w.r.t. the canonical angle) initial conditions on the initial invariant torus is studied in the section~\ref{EnEvol}. In the same section the exact expressions for the energy moments of arbitrary order taken at an arbitrary moment of time are analytically derived and the corresponding characteristic function is analytically constructed in the form of infinite series. The limit of small initial ensemble's energy is considered in section~\ref{SmallE}. A simple formula coinciding with the one obtained in~\cite{LastLetter} for the energy distribution function is analytically derived in this limit. In section~\ref{Num} the energy distribution characteristic function is numericaly evaluated for certain values of the system's parameters. Also, the energy distribution function is numerically calculated in some particular cases. Two appendices are added for the sake of the selfconsistency of the paper.

We also would like to note that for the quantized time-dependent harmonic oscillator with an external forcing the time evolution of the displacement parameter can be described in terms of the classical effective Hamiltonian having the same form as the one studied in the rest of this paper. The particular implementation of the quantized time-dependent harmonic oscillator with the external forcing is the key model exploited in geometrical ion trap quantum computations~\cite{GeomTRIO,PRA3}.

\section{Equations of motion and their solution using WKB approach}\label{Eq}

We consider the time-dependent harmonic oscillator with the external time-dependent forcing described by the Hamiltonian
\begin{equation}\label{H}
  H = \frac{1}{2} p^2 + \frac{1}{2} \omega^2 (t) q^2 + \tilde{\gamma} q f(t),
\end{equation}
where $q$ and $p$ are canonically conjugate generalized coordinate and momentum, $\omega (t)$ is the time-dependent oscillator frequency, $f(t)$ is the time-dependent external force and $\tilde{\gamma}$ is the coupling strength. Corresponding Hamilton equations of motion are
\begin{eqnarray}\label{HamEq}
\dot{q} = \frac{\partial H}{\partial p} = p , \nonumber \\
\dot{p} = - \frac{\partial H}{\partial q} = - \omega^2 (t) q - \tilde{\gamma} f(t) .
\end{eqnarray}
Here the dot denotes the time derivative. This system of the first order differential equations is equivalent to the single second order differential equation:
\begin{equation}\label{NonHomEq}
  \ddot{q} + \omega^2 (t) q = - \tilde{\gamma} f(t).
\end{equation}
This is a nonhomogeneous ordinary differential equation and it is known that its general solution is the sum of the general solution of the corresponding homogeneous equation and some particular solution of the nonhomogeneous equation~\cite{Korn}. The exact solution of the homogeneous equation in terms of WKB series can be found in the Refs.~\cite{RobnikRomanovski,WKBMod} and the summary is given in~\ref{AppA}. The general solution of the nonhomogeneous equation~(\ref{NonHomEq}) can be expressed in terms of Green function~\cite{Korn}. Namely,
\begin{eqnarray}\label{q}
  q (\lambda) = q_h (\lambda) + f_q =  w_+ \exp{\left\{ \frac{1}{\epsilon} \sigma_+ (\lambda) \right\}} + w_- \exp{\left\{ \frac{1}{\epsilon} \sigma_- (\lambda) \right\}} \nonumber \\ 
- \tilde{\gamma} \int\limits_{\lambda_0}^{\lambda_f} G (\lambda , x) f(x) dx, \quad \lambda_0 < \lambda < \lambda_f ,
\end{eqnarray}
where the notations are explained in~\ref{AppA} and $f_q$ is introduced for further convenience. The Green function is given by the expression~\cite[9.3-9]{Korn}:
\begin{equation}\label{GreenF}
  G (\lambda , x) = - \frac{q_+ (\lambda) q_- (x) - q_- (\lambda) q_+ (x)}{q_+ (x) q^\prime_- (x) - q_- (x) q^\prime_+ (x)} U(\lambda - x).
\end{equation}
Here $q_+$ and $q_-$ denote two linearly independent solutions of the homogeneous equation:
\begin{equation}\label{qq}
  q_+ = \exp{\left\{ \frac{1}{\epsilon} \sigma_+ \right\}}, \quad q_- = \exp{\left\{ \frac{1}{\epsilon} \sigma_- \right\}}
\end{equation}
and $U(z)$ is the unit step function defined as
\begin{equation}\label{Step}
  U(z) = \left\{
\begin{array}{c}
  0, \quad z<0, \\
  1, \quad z>0 .
\end{array}
\right.
\end{equation}
The denominator in~(\ref{GreenF}) is just the Wronskian of the relevant equation and is constant for all $x$.
Using the results of the~\ref{AppA} the expression~(\ref{GreenF}) can be reduced to
\begin{equation}\label{Green1}
  G(\lambda , x) = \frac{\epsilon}{B(x)} \exp{\left\{ \frac{r(\lambda , x)}{\epsilon} \right\}} \sin{\left(\frac{s(\lambda , x)}{\epsilon}\right)} U(\lambda - x), 
\end{equation}
where
\begin{equation}\label{rs}
  r (\lambda , x) = \int\limits_x^\lambda A(z) dz , \quad s(\lambda ,x) = \int\limits_x^\lambda B(z) dz .
\end{equation}
The requirement of the phase space volume conservation leads to the relation
\begin{equation}\label{Conserv}
  \frac{\exp{\left(\frac{r(\lambda , x)}{\epsilon}\right)}}{B(x)} = \frac{1}{\sqrt{B(\lambda) B(x)}}.
\end{equation}
Thus the Green function can be represented in the form:
\begin{equation}\label{Green2}
  G(\lambda , x) = \frac{\epsilon}{\sqrt{B(\lambda) B(x)}} \sin{\left(\frac{s(\lambda,x)}{\epsilon}\right)} U(\lambda -x).
\end{equation} 
The momentum evolution can be determined using the relationship $p(\lambda) = \epsilon q^\prime (\lambda)$ and it is given by the expression
\begin{eqnarray}\label{p}
  p (\lambda) = w_+ \sigma_+^\prime (\lambda) \exp{\left\{ \frac{1}{\epsilon} \sigma_+ (\lambda) \right\}} + w_- \sigma_-^\prime (\lambda) \exp{\left\{ \frac{1}{\epsilon} \sigma_- (\lambda) \right\}} \nonumber \\ - \epsilon \tilde{\gamma} \int\limits_{\lambda_0}^{\lambda_f} G^\prime_\lambda (\lambda , x) f(x) dx = p_h (\lambda) + f_p,
\end{eqnarray} 
where $p_h (\lambda)$ is the solution of the equations of motion when the external force is zero ($f(t) \equiv 0$) and $f_p$ is proportional to the integral of Green function's derivative multiplied by the force.

\section{Energy distribution moments and characteristic function}\label{EnEvol}

In this section we analytically derive the exact expressions for the energy distribution moments taken at arbitrary time $t = t_1$ for the ensemble of uniformly distributed (w.r.t. the canonical angle) initial conditions on the initial invariant torus. Using them we construct the characteristic function of the energy distribution in form of an infinite power series. First, we note that the generalized coordinate and momentum of the system at time $t_1$ are the linear combinations of the same quantities taken at the initial time moment $t=t_0$ plus nonhomogeneous terms originating from the forcing. Namely,
\begin{equation}\label{NonHomMap}
  \left(
\begin{array}{c}
  q_1 \\
  p_1
\end{array} \right) = 
\left(
\begin{array}{cc}
  a & b \\
  c & d
\end{array} \right)
\left(
\begin{array}{c}
  q_0 \\
  p_0
\end{array} \right)
+ \left(
\begin{array}{c}
  f_q \\
  f_p
\end{array} \right),
\end{equation}
where $q_1 = q(t_1)$, $p_1 = p(t_1)$, $q_0 = q(t_0)$, $p_0 = p(t_0)$, the coefficients $a$, $b$, $c$ and $d$ are the same as in the homogeneous case and the corresponding expressions can be found in the~\ref{AppA}. The quantities $f_q$ and $f_p$ are defined in~(\ref{q}) and (\ref{p}) and are given by the expressions
\begin{eqnarray}\label{fqfp}
  f_q = - \tilde{\gamma} \int\limits_{\lambda_0}^{\lambda_f} G(\lambda , x) f(x) dx , \nonumber \\
  f_p = - \epsilon \tilde{\gamma} \int\limits_{\lambda_0}^{\lambda_f} G^\prime_\lambda (\lambda , x) f(x) dx .
\end{eqnarray}
The energy of the system $E_1 = H(q(t_1), p(t_1), t_1)$ at the time $t=t_1$ is 
\begin{equation}\label{E1}
  E_1 = \frac{1}{2} p_1^2 + \frac{1}{2} \omega_1^2 q_1^2 + \tilde{\gamma} q_1 f_1,
\end{equation} 
where $\omega_1 = \omega (t_1)$ and $f_1 = f(t_1)$. If the initial energy of the ensemble equals $E_0 = H (q(t_0), p(t_0), t_0)$ then using~(\ref{NonHomMap}), assuming $f_0 = 0$ (without losing generality) and performing the transformation
\begin{equation}\label{q0p0}
  q_0 = \sqrt{\frac{2E_0}{\omega_0^2}} \cos{\phi} , \quad p_0 = \sqrt{2E_0} \sin{\phi}
\end{equation}
with $\omega_0 = \omega(t_0)$, one obtains
\begin{equation}\label{E1alpha}
  E_1 = \alpha \cos^2{\phi} + \beta \sin^2{\phi} + \gamma \sin{2 \phi} + \sigma \cos{\phi} + \rho \sin{\phi} + \delta .
\end{equation}
Here
\begin{eqnarray}\label{greeks}
  \alpha = \frac{E_0}{\omega^2_0} \left( c^2 + \omega_1^2 a^2 \right), \nonumber \\
  \beta = E_0 \left( d^2 + \omega_1^2 b^2 \right) , \nonumber \\
  \gamma = \frac{E_0}{\omega_0} \left( cd + \omega_1^2 ab \right) , \nonumber \\
  \sigma = \sqrt{\frac{2E_0}{\omega_0^2}} \left( c f_p + \omega_1^2 a f_q + \tilde{\gamma} a f_1 \right),\nonumber \\
  \rho = \sqrt{2E_0} \left( d f_p + \omega^2_1 b f_q + \tilde{\gamma} b f_1 \right) , \nonumber \\
  \delta = \frac{1}{2} f_p^2 + \frac{1}{2} \omega_1^2 f_q^2 + \tilde{\gamma} f_q f_1 .
\end{eqnarray}
Using the equation~(\ref{E1alpha}) one can calculate the average energy of the ensemble at time $t_1$:
\begin{equation}\label{aveE1}
  \bar{E_1} = \frac{1}{2} \left( \alpha + \beta \right) + \delta  
\end{equation}
and the deviation of the energy from its mean value
\begin{equation}\label{DevE1}
  E_1 - \bar{E_1} = \sqrt{\kappa^2 + \gamma^2} \sin{\left( 2\phi + \psi_1 \right)} + \sqrt{\sigma^2 + \rho^2} \sin{\left( \phi + \psi_2  \right)} ,
\end{equation}
where 
\begin{equation}\label{kappa}
  \kappa = \frac{1}{2} \left( \alpha - \beta \right) , \quad
  \tan{\psi_1} = \frac{\kappa}{\gamma} , \quad
  \tan{\psi_2} = \frac{\sigma}{\rho} .
\end{equation}
Using equation~(\ref{DevE1}) we obtain the exact expressions for the even and odd moments of the energy distribution. The main steps of a bit bulky derivation are given in the~\ref{AppB}. The results are
\begin{eqnarray}\label{evenMoments}
  \overline{\left( E_1 - \bar{E_1} \right)^{2n}} \equiv \frac{1}{2\pi} \int\limits_{0}^{2\pi} d \phi \left( E_1 - \bar{E_1} \right)^{2n}= \frac{(2n-1)!!}{2^n n!} \left( \tilde{A}^{2n} + \tilde{B}^{2n} \right) \nonumber \\
  + \sum\limits_{p=1}^n \sum\limits_{q=0}^{n-p} \sum\limits_{s=0}^{p-1} \frac{(-1)^{n+q+s-1}}{2^{2n-1}} \tilde{A}^{2n-2p+1} \tilde{B}^{2p-1} \left(\begin{array}{c} 2n \\ 2p-1 \end{array} \right) \left( \begin{array}{c} 2n-2p+1 \\ q \end{array} \right) \nonumber \\ 
\times \left( \begin{array}{c} 2p-1 \\ s \end{array} \right) \cos{\left[ \left( 2n - 2p -2q +1 \right)\psi_1 - \left( 2p - 2s -1 \right)\psi_2 \right]} \nonumber \\
\times \delta_{2\left( 2n - 2p -2q +1 \right), 2p-2s-1} + \sum\limits_{p=1}^{n-1} \sum\limits_{q=0}^{n-p-1} \sum\limits_{s=0}^{p-1} \frac{(-1)^{n-q-s}}{2^{2n-1}} \tilde{A}^{2n-2p} \tilde{B}^{2p} \nonumber \\ 
\times \left( \begin{array}{c} 2n \\ 2p \end{array} \right) \left( \begin{array}{c} 2(n-p) \\ q \end{array} \right) \left( \begin{array}{c} 2p \\ s \end{array} \right) \cos{\left[ 2\left( n-p-q  \right)\psi_1 - 2\left( p-s \right)\psi_2 \right]} \nonumber \\
\times \delta_{2(n-p-q), p-s} + \sum\limits_{p=1}^{n-1} \frac{1}{2^{2n}} \tilde{A}^{2n-2p} \tilde{B}^{2p} \left( \begin{array}{c} 2n \\ 2p \end{array} \right) \left( \begin{array}{c} 2(n-p) \\ n-p \end{array} \right) \left( \begin{array}{c} 2p \\ p \end{array} \right),
\end{eqnarray}
\begin{eqnarray}\label{oddMoments}
  \overline{\left( E_1 - \bar{E_1} \right)^{2n-1}} = \sum\limits_{p=1}^{n-1} \sum\limits_{q=0}^{n-p-1} \sum\limits_{s=0}^{p-1} \frac{(-1)^{n+s-q}}{2^{2n-2}} \tilde{A}^{2n-2p} \tilde{B}^{2p-1} \left( \begin{array}{c} 2n-1 \\ 2p-1 \end{array} \right) \nonumber \\
\times \left( \begin{array}{c} 2(n-p) \\ q \end{array} \right) \left( \begin{array}{c} 2p-1 \\ s \end{array} \right) \sin{\left[ 2\left(n-p-q \right)\psi_1 - \left( 2p-2s-1 \right)\psi_2 \right]} \nonumber \\
\times \delta_{4(n-p-q), 2p-2s-1} + \sum\limits_{p=1}^{n-1} \sum\limits_{q=0}^{n-p-1} \sum\limits_{s=0}^{p-1} \frac{(-1)^{n+q-s-1}}{2^{2n-2}} \tilde{A}^{2n-2p-1} \tilde{B}^{2p} \nonumber \\
\times \left( \begin{array}{c} 2n-1 \\ 2p \end{array} \right) \left( \begin{array}{c} 2(n-p) -1 \\ q \end{array} \right) \left( \begin{array}{c} 2p \\s \end{array} \right) \nonumber \\
\times \sin{\left[ \left( 2n -2p -2q -1 \right) \psi_1 - 2\left( p-s \right)\psi_2 \right]}  \delta_{2n-2p-2q-1, p-s}.
\end{eqnarray}
Here
\begin{equation}\label{mainAB}
  \tilde{A} = \sqrt{\kappa^2 + \gamma^2} , \quad \tilde{B} = \sqrt{\sigma^2 + \rho^2}.
\end{equation}
For any specific numerical value of $n$ the sums in~(\ref{evenMoments})-(\ref{oddMoments}) can be analytically evaluated using one of the algebraic software packages, for example {\it Mathematica}. Below we provide the expressions for the few lowest moments:
\begin{eqnarray}\label{lowestMoments}
  \overline{\left( E_1 - \bar{E_1} \right)^2} \equiv \mu^2 = \frac{1}{2}\left(\tilde{A}^2 + \tilde{B}^2 \right), \nonumber \\
  \overline{\left( E_1 - \bar{E_1} \right)^3} = - \frac{3}{4} \tilde{A} \tilde{B}^2 \sin{\left( \psi_1 - 2\psi_2 \right)}, \nonumber \\
  \overline{\left( E_1 - \bar{E_1} \right)^4} = \frac{3}{8}\left( \tilde{A}^4 + 4 \tilde{A}^2 \tilde{B}^2 + \tilde{B}^4  \right).
\end{eqnarray}
The first moment $\overline{\left( E_1 - \bar{E_1}  \right)} = 0$ as it should be. Equations~(\ref{evenMoments})-(\ref{oddMoments}) allow us to construct the characteristic function of the energy distribution in the form of an infinite series according to the following formula:
\begin{eqnarray}\label{CharFunc}
  f (y) = 1 + \sum\limits_{m=1}^{\infty} \frac{i^{2m}}{(2m)!}\overline{\left(E_1 - \bar{E_1} \right)^{2m}} y^{2m} \nonumber \\
 + \sum\limits_{m=1}^\infty \frac{i^{2m-1}}{(2m-1)!} \overline{\left( E_1 - \bar{E_1} \right)^{2m-1}} y^{2m-1}.
\end{eqnarray}
It is seen that the characteristic function is complex. It means that the energy distribution is not an even function of $E_1 - \bar{E_1}$ in contrast to the case of the time-dependent harmonic oscillator without forcing~\cite{Robnik2006,RobnikRomanovski,LastLetter,WKBMod}. The same conclusion follows from the equation~(\ref{oddMoments}), since the odd moments are non-zero in the general case.

\section{Small initial energy limit}\label{SmallE}

The expressions~(\ref{evenMoments})-(\ref{oddMoments}) can be significantly simplified if we consider the evolution of the initial ensemble distributed close enough to the point $(q=0, p=0)$ and thus having small initial energy $E_0$. In this case the series for the characteristic function~(\ref{CharFunc}) can be summed up and the expression for the distribution function can be analytically obtained. From equations~(\ref{greeks}), (\ref{mainAB}) it follows that $\tilde{A} \sim E_0$ and $\tilde{B} \sim \xi E_0^{1/2}$, where $\xi$ has the dimension of the square root of the energy, but it does not depend on $E_0$. Therefore when $E_0$ is small, $\tilde{A} \ll \tilde{B}$. Starting from either~(\ref{DevE1}) or directly from~(\ref{evenMoments})-(\ref{oddMoments}) it follows that the even moments in this case are the same as for the time-dependent harmonic oscillator without forcing considered in the Refs.~\cite{Robnik2006,RobnikRomanovski,LastLetter} and are given by the expression
\begin{equation}\label{S:moments}
  \overline{\left( E_1 - \bar{E_1} \right)^{2n}} \approx \frac{(2n-1)!!}{n!} \mu^{2n},
\end{equation} 
where $\mu^2 \approx \tilde{B}^2 /2$ is the second moment or variance. The odd moments are zero in this approximation. The average energy value equals $\bar{E_1} \approx \delta$ and has also the meaning of the energy acquired by the system at the time $t_1$ starting its motion at the initial point $(q_0 = 0, p_0 = 0)$. The characteristic function in this case is given by~\cite{LastLetter}
\begin{equation}\label{S:CharFun}
  f (y) \approx \sum\limits_{m=0}^\infty \frac{1}{(m!)^2}  \left( - \frac{\mu^2 y^2}{2}  \right)^m = J_0 \left( \mu y \sqrt{2} \right),
\end{equation}
where $J_0 (z)$ is the zero order Bessel function. It yields the following distribution function~\cite{LastLetter}:
\begin{equation}\label{S:dist}
  P(E_1) = \frac{1}{\pi \sqrt{2\mu^2 - \left(E_1 - \bar{E_1}\right)^2}} .
\end{equation}
This function has two singularities at $\left( E_1 - \bar{E_1} \right) = \pm \mu \sqrt{2}$ and it is even with respect to $\left(E_1 - \bar{E_1} \right)$.

\section{Numerical results}\label{Num}

\begin{figure}
\epsfxsize=0.5\linewidth
\epsfbox{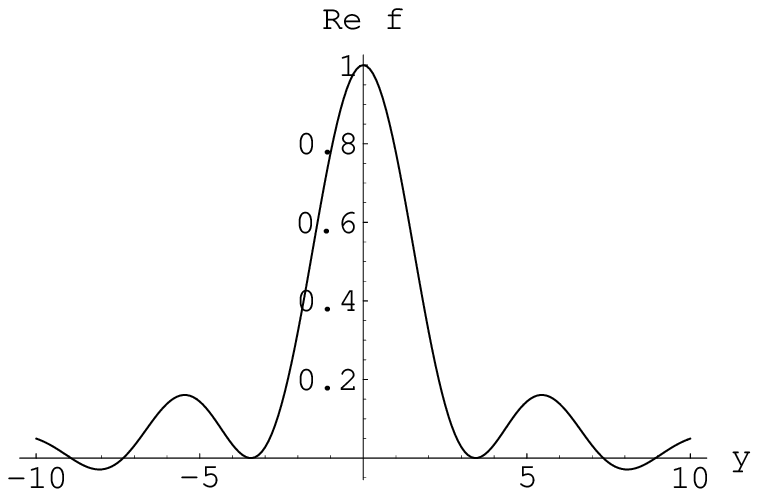}
\epsfxsize=0.5\linewidth
\epsfbox{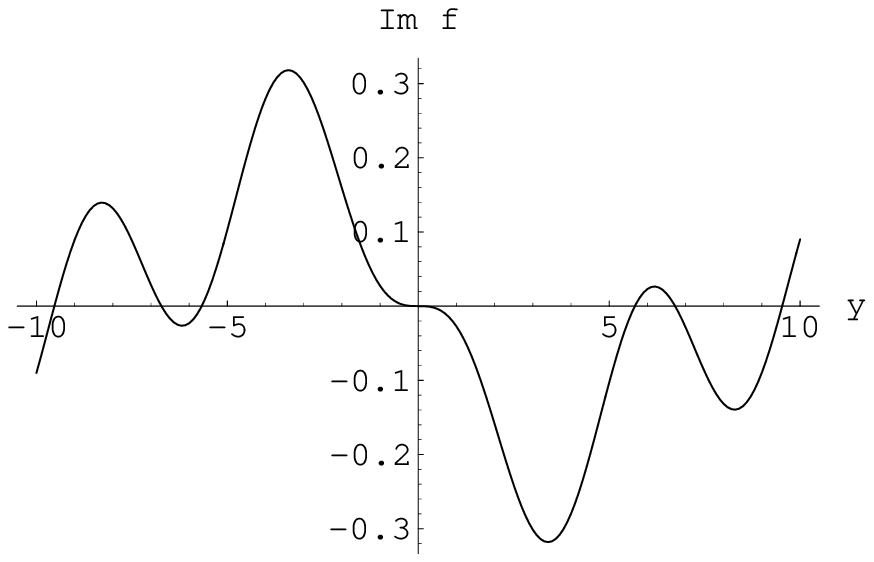}
\caption{The plots of the real and imaginary parts of the energy distribution characteristic function for the certain values of the parameters. Namely, $\tilde{A}=0.7$, $\tilde{B}=0.7$, $\psi_1 = \pi /4$ and $\psi_2 = \pi /4$.}\label{fig:ChF}
\end{figure}

The characteristic function in~(\ref{CharFunc}) can be numerically evaluated for certain values of the parameters. The plots of its real and imaginary parts, when $\tilde{A}=0.7$, $\tilde{B}=0.7$, $\psi_1 = \pi /4$ and $\psi_2 = \pi /4$, are presented in figure~\ref{fig:ChF}. 
Here the first forty energy distribution moments have been summed up in order to approximate the characteristic function. We also extended the summation up to the sixty moments (not shown) in order to convince ourselves that the result does not change. 

\begin{figure}
 \centering
 \epsfxsize=0.5\linewidth
 \epsfbox{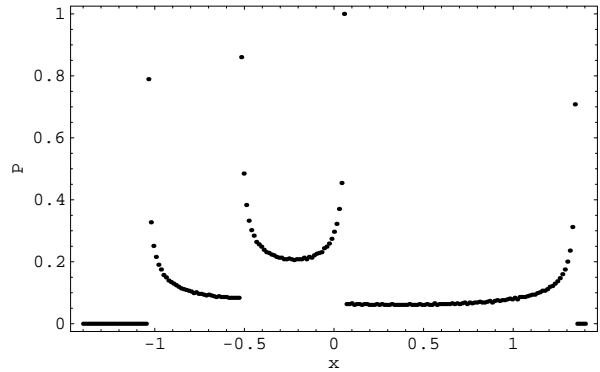}
 \caption{The normalized histogram approximating the energy distribution function $P(x)$, when  $\tilde{A}=0.7$, $\tilde{B}=0.7$, $\psi_1 = \pi/4$ and $\psi_2 = \pi/4$.}\label{fig:Eq}
\end{figure}

\begin{figure}
  \centering
  \epsfxsize=0.5\linewidth
  \epsfbox{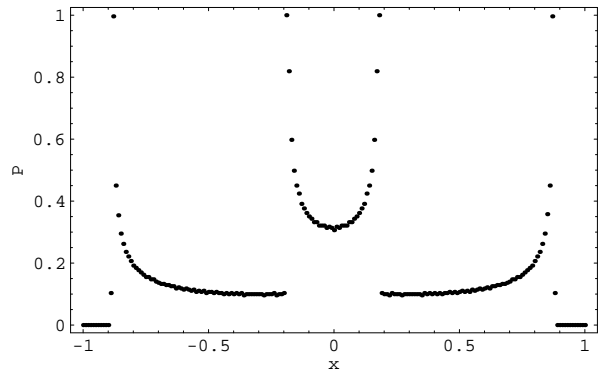}
  \caption{The normalized histogram approximating the energy distribution function $P(x)$, when  $\tilde{A}=0.5$, $\tilde{B}=0.5$, $\psi_1 = \pi/2$ and $\psi_2 = \pi/4$. The distribution function is symmetric with respect to $x$.}\label{fig:3}
\end{figure}

\begin{figure}
\centering
\epsfxsize=0.5\linewidth
\epsfbox{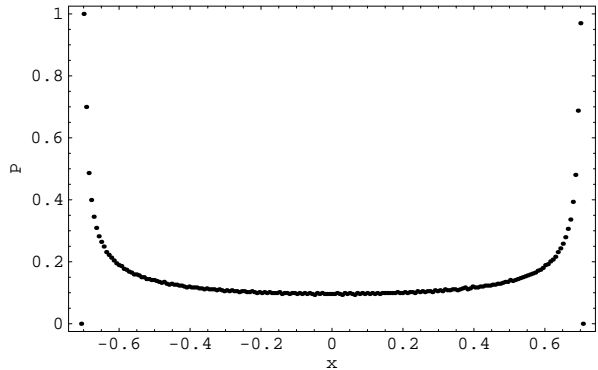}
\caption{The normalized histogram approximating the energy distribution function $P(x)$ in the small initial energy limit. Parameters values are $\tilde{A}=0.007$, $\tilde{B}=0.7$, $\psi_1 = \pi/3$ and $\psi_2 = \pi/4$. The distribution function has only two singularities.}\label{fig:4}
\end{figure}

Now we numerically obtain the energy distribution function. For this purpose we exploit equation~(\ref{DevE1}). We divide the range from $0$ to $2\pi$ in $10 000$ intervals, let $\phi$ have a value in each of this small intervals and find how many corresponding values of $x \equiv \left( E_1 - \bar{E_1} \right)$ fall into one of the $200$ intervals covering the range from $-\left(\tilde{A}+ \tilde{B} \right)$ to $\tilde{A} + \tilde{B}$. After the normalization we obtain the histogram approximating the energy distribution function. We perform the corresponding computations for several values of the parameters. Primarily, we considered the case when $\tilde{A}=0.7$, $\tilde{B}=0.7$, $\psi_1 = \pi/4$ and $\psi_2 = \pi/4$ (the same values as used in the figure~\ref{fig:ChF}). The result is shown in figure~\ref{fig:Eq}.
It is seen that the distribution function has four singularities in this case instead of two reported for the time-dependent harmonic oscillator without external forcing~\cite{Robnik2006,RobnikRomanovski,LastLetter} or obtained in the small initial energy limit in section~\ref{SmallE}. Two additional singularities appeared due to the additional term in~(\ref{DevE1}) proportional to $\sin{\left(\phi + \psi_2\right)}$. Also the distribution function is not even with respect to $x$ as expected. These conclusions are supported by the numerical calculations at other values of the parameters (not shown). However, we found that the distribution function can be a symmetric one with respect to $x$ for some particular parameter values. For example, this is the case if $\tilde{A} = 0.5$, $\tilde{B}=0.5$, $\psi_1 = \pi /2$ and $\psi_2 = \pi /4$, where according to~(\ref{lowestMoments}) the third moment is zero. The result of the computation is shown in figure~\ref{fig:3}.
Indeed, it is clear from~(\ref{DevE1}) that $P(x)$ is an even function of $x$ if $\psi_1 = 2\psi_2$ in which case all odd moments~(\ref{oddMoments}) vanish, which has been confirmed using {\it Mathematica} up to and including 79th moment.
We have also checked the behaviour of the distribution function in the small initial energy limit. The corresponding plot is given in figure~\ref{fig:4}.
The following parameters values were chosen: $\tilde{A}=0.007$, $\tilde{B}=0.7$, $\psi_1 = \pi/3$ and $\psi_2 = \pi/4$. It is seen that the energy distribution function has only two singularities in this limit as it should be according to the equation~(\ref{S:dist}) and it is symmetric w.r.t. $x$.


\section{Conclusions}

We have constructed the general solution of the equations of motion for the time-dependent harmonic oscillator with external forcing. The approach based on WKB method developed in~\cite{Robnik2006,RobnikRomanovski,LastLetter,WKBMod} was exploited for this purpose. Using it we have analytically derived the exact expressions for the moments of the energy distribution of arbitrary order calculated at arbitrary moment of time for the case of the ensemble of uniformly distributed (w.r.t. cannonical angle) initial conditions on the invariant torus. We have obtained the characteristic function of the energy distribution in the form of an infinite series. It appears to be a complex one, since the odd moments are non-zero in distinction to the case of the time-dependent harmonic oscillator without forcing considered in~\cite{Robnik2006,RobnikRomanovski,LastLetter}. The energy distribution function is not even w.r.t. $x = \left(E_1 - \bar{E_1} \right)$ in the presense of external forcing. We have considered the limit of the small initial ensemble's energy and analytically obtained the expressions for both the characteristic function and the energy distribution function in this case. They coincide with the expressions for the same quantities in the case of the time-dependent harmonic oscillator without forcing but the physical meaning and origin of the parameters is now different. The characteristic function is real and the energy distribution is an even function w.r.t. $x = (E_1 - \bar{E_1})$ and has two singularities in this limit. We have numerically obtained the characteristic function and the energy distribution for some particular values of the parameters for the general case, where the energy distribution function has four singularities, but in the small initial energy limit their number reduces to two, indeed. The results obtained can be useful in the context of geometric ion trap quantum computations~\cite{GeomTRIO,PRA3} where the particular implementation of the investigated model of the time-dependent harmonic oscillator with external forcing plays the crucial role. 

\section{Acknowledgements}

A.V.K. gratefully acknowledges the warm hospitality by Prof. Marko Robnik at the CAMTP --- Center for Applied Mathematics and Theoretical Physics, University of Maribor, and the financial support by Ad Futura foundation.

\clearpage

\appendix

\section{Solution of the homogeneous equations of motion in the framework of WKB approach}\label{AppA}

In this section we give a brief summary of the results concerning the solution of the homogeneous equation~(\ref{NonHomEq}), which is
\begin{equation}\label{HomEq}
  \ddot{q} + \omega^2 (t) q = 0.
\end{equation}
More details can be found in~\cite{Robnik2006,RobnikRomanovski}. Let us first introduce the dimensionless time $\lambda = \epsilon t$, where $\epsilon = 1/T$ and $T$ is the typical time scale under consideration of the system's dynamics. Equation~(\ref{HomEq}) is transformed to
\begin{equation}\label{ReHomEq}
  \epsilon^2 q^{\prime \prime} (\lambda) + \omega^2 (\lambda) q (\lambda) = 0.
\end{equation}
The prime denotes the differentiation with respect to $\lambda$. The solution of the equation~(\ref{ReHomEq}) is sought in the form
\begin{equation}\label{HomSolForm}
  q (\lambda) = w \exp{\left\{ \frac{1}{\epsilon} \sigma (\lambda) \right\}}.
\end{equation}
Here $w$ is some constant with the dimension of length. After the substitution one gets:
\begin{equation}\label{SigEq}
  \left( \sigma^{\prime} (\lambda) \right)^2 + \epsilon \sigma^{\prime \prime} (\lambda) = - \omega^2 (\lambda).
\end{equation}
The WKB expansion for the $\sigma$ is
\begin{equation}\label{SigExp}
  \sigma (\lambda) = \sum\limits_{k=0}^\infty \epsilon^k \sigma_k (\lambda).
\end{equation}
The substitution of~(\ref{SigExp}) into (\ref{SigEq}) and the comparision of the coefficients at the same powers of $\epsilon$ gives the following recursion relation:
\begin{eqnarray}\label{Recurs}
  \sigma_0^{\prime 2} = - \omega^2 (\lambda) ; \quad \sigma_1^\prime = - \frac{\sigma_0^{\prime \prime}}{2\sigma_0^\prime} ; \nonumber \\ \quad \sigma_n^\prime = - \frac{1}{2 \sigma_0^\prime} \left( \sum\limits_{k=1}^{n-1} \sigma_k^\prime \sigma_{n-k}^\prime + \sigma_{n-1}^{\prime \prime} \right), \quad n\ge 2.
\end{eqnarray} 
Equation~(\ref{ReHomEq}) has two linearly independent solutions given by the choice of $\sigma_{0,+}^\prime (\lambda) = i \omega (\lambda)$ or $\sigma_{0,-}^\prime (\lambda) = - i \omega (\lambda)$ as the starting term in the recursion~(\ref{Recurs}). Since the frequency $\omega (\lambda)$ is real, it follows that all functions $\sigma_{2k+1}^\prime$ are real and all functions $\sigma_{2k}^\prime$ are pure imaginary and the following relations hold:
\begin{equation}\label{SymRel}
  \sigma_{2k,+}^\prime = - \sigma_{2k,-}^\prime , \quad \sigma_{2k+1,+}^\prime = \sigma_{2k+1,-}^\prime = \sigma_{2k+1}^\prime ,
\end{equation}
where $k = 0,1,2,\ldots$. Thus
\begin{equation}\label{SAB}
  \sigma_+^\prime (\lambda) = A (\lambda) + i B(\lambda), \quad \sigma_-^\prime (\lambda) = A (\lambda) - i B (\lambda),
\end{equation}
where
\begin{equation}\label{AB}
  A (\lambda) = \sum\limits_{k=0}^\infty \epsilon^{2k+1} \sigma_{2k+1}^\prime (\lambda) , \quad B (\lambda) = - i \sum\limits_{k=0}^\infty \epsilon^{2k} \sigma_{2k,+}^\prime (\lambda)
\end{equation}
are both real quantities. The integration yields
\begin{equation}\label{Srs}
  \sigma_+ (\lambda) = r(\lambda) + i s(\lambda), \quad \sigma_- (\lambda) = r (\lambda) - i s(\lambda) , 
\end{equation}
where
\begin{equation}\label{rs}
  r (\lambda) = \int\limits_{\lambda_0}^{\lambda} A (z) dz , \quad s (\lambda) = \int\limits_{\lambda_0}^\lambda B(z) dz .
\end{equation}
Here $\lambda_0$ corresponds to the initial moment of time. Thus the general solution of the homogeneous equation~(\ref{ReHomEq}) is given by
\begin{equation}\label{HomSol}
  q_h (\lambda) = w_+ \exp{\left\{ \frac{1}{\epsilon} \sigma_+ (\lambda) \right\}} + w_- \exp{\left\{ \frac{1}{\epsilon} \sigma_- (\lambda) \right\}}.
\end{equation}

The initial values of the coordinate and momentum of the homogeneous system ($f (t) \equiv 0$), namely, $q_0 = q(t_0)$ and $p_0 = p(t_0)$ are connected with their final values $q_1 = q (t_1)$ and $p_1 = p(t_1)$ at time $t=t_1$ by the linear transition map:
\begin{equation}\label{FConn}
  \Phi : \left(
\begin{array}{c}
q_0 \\
p_0
\end{array}
 \right) \rightarrow
\left(
\begin{array}{c}
q_1 \\
p_1
\end{array}
\right),
\end{equation}  
where
\begin{equation}\label{F}
  \Phi = \left(
\begin{array}{cc}
a & b \\
c & d
\end{array}
\right)
\end{equation}
and the matrix elements are given by the expressions~\cite{Robnik2006,RobnikRomanovski}:
\begin{eqnarray}\label{abcd}
  a  = - \frac{1}{\sqrt{B(\lambda_0)B(\lambda_1)}} \left[ A(\lambda_0) \sin{\left( \frac{s(\lambda_1)}{\epsilon}\right)} - B(\lambda_0) \cos{\left( \frac{s(\lambda_1)}{\epsilon} \right)} \right], \nonumber \\
b = \frac{1}{\sqrt{B(\lambda_0)B(\lambda_1)}} \sin{\left( \frac{s(\lambda_1)}{\epsilon} \right)}, \nonumber \\
c = - \frac{1}{\sqrt{B(\lambda_0)B(\lambda_1)}} \left[ \left(A(\lambda_0) A(\lambda_1) + B (\lambda_0) B(\lambda_1)\right) \sin{\left( \frac{s(\lambda_1)}{\epsilon}\right)} \right. \nonumber \\ 
\left. +\left(A(\lambda_0) B(\lambda_1) - A(\lambda_1) B (\lambda_0) \right) \cos{\left(\frac{s(\lambda_1)}{\epsilon}\right)} \right], \nonumber \\
d = \frac{1}{\sqrt{B(\lambda_0) B(\lambda_1)}} \left[ A(\lambda_1) \sin{\left(\frac{s(\lambda_1)}{\epsilon}\right)} + B(\lambda_1) \cos{\left(\frac{s(\lambda_1)}{\epsilon}\right)} \right].
\end{eqnarray}

\section{Derivation of the exact expressions for the energy moments}\label{AppB}

In this section we provide the main steps of the energy distribution moments derivation. Two cases have to be considered. Namely, the case of the even moments and the case of the odd ones. We start with the even moments of arbitrary order. They are
\begin{eqnarray}\label{B1}
  \overline{\left( E_1 - \bar{E_1} \right)^{2n}} = \frac{1}{2\pi} \int\limits_{0}^{2\pi} d \phi \left( E_1 - \bar{E_1} \right)^{2n} \nonumber \\
= \frac{1}{2\pi} \int\limits_{0}^{2\pi} d \phi \left\{ \tilde{A} \sin{\left( 2\phi + \psi_1 \right)} + \tilde{B} \sin{\left( \phi + \psi_2 \right)} \right\}^{2n},
\end{eqnarray}
where the equation~(\ref{DevE1}) was used, and $\tilde{A} = \sqrt{\kappa^2 + \gamma^2}$ and $\tilde{B} = \sqrt{\sigma^2 + \rho^2}$. Applying Newton's binomial formula to~(\ref{B1}) and expanding the powers of sines~\cite[1.320]{GR} the even moments can be represented in the form:
\begin{eqnarray}\label{B2}
  \overline{\left( E_1 - \bar{E_1} \right)^{2n}} = \frac{1}{2\pi} \tilde{A}^{2n} \int\limits_{0}^{2\pi} d \phi \sin^{2n}{\left( 2 \phi + \psi_1 \right)} \nonumber \\ 
+ \frac{1}{2\pi} \sum\limits_{p=1}^n \tilde{A}^{2n-2p+1} \tilde{B}^{2p-1} \left( \begin{array}{c} 2n \\ 2p-1 \end{array} \right) \int\limits_0^{2\pi} d \phi \left\{ \frac{1}{2^{2n-2p}} \sum\limits_{q=0}^{n-p} (-1)^{n-p+q} \right. \nonumber \\
\times \left. \left( \begin{array}{c} 2n-2p+1 \\ q \end{array} \right) \sin{\left( \left[ 2n-2p-2q+1 \right] \left[ 2 \phi + \psi_1 \right]  \right)}   \right\} \nonumber \\
\times\left\{ \frac{1}{2^{2p-2}} \sum\limits_{s=0}^{p-1} (-1)^{p+s-1} \left( \begin{array}{c} 2p-1 \\ s \end{array} \right) \sin{\left( \left[ 2p-2s-1 \right] \left[ \phi + \psi_2 \right]  \right)} \right\} \nonumber \\
+ \frac{1}{2\pi} \sum\limits_{p=1}^{n-1} \tilde{A}^{2n-2p} \tilde{B}^{2p} \left( \begin{array}{c} 2n \\ 2p \end{array} \right) \int_0^{2\pi} d \phi \left\{ \frac{1}{2^{2(n-p)}} \left[ \sum\limits_{q=0}^{n-p-1} (-1)^{n-p-q} 2 \left( \begin{array}{c} 2(n-p) \\ q \end{array}   \right) \right. \right. \nonumber \\
\left. \left. \times \cos{\left( 2 \left[ n-p-q \right] \left[ 2 \phi + \psi_1 \right] \right)} + \left( \begin{array}{c} 2 (n-p) \\ n-p \end{array} \right)  \right]  \right\} \nonumber \\
\times \left\{ \frac{1}{2^{2p}} \left[ \sum\limits_{s=0}^{p-1} (-1)^{p-s} 2 \left( \begin{array}{c} 2p \\s \end{array} \right) \cos{\left( 2 \left[ p-s \right] \left[ \phi +\psi_2 \right] \right)} + \left( \begin{array}{c} 2p \\ p \end{array}  \right)  \right]  \right\} \nonumber \\
+ \frac{1}{2\pi} \tilde{B}^{2n} \int_0^{2\pi} d \phi \sin^{2n}{\left( \phi + \psi_2 \right)}.
\end{eqnarray}  
The integration over the angle $\phi$ yields the result (see~\cite[2.511]{GR}):
\begin{eqnarray}\label{Beven}
  \overline{\left( E_1 - \bar{E_1} \right)^{2n}} \equiv \frac{1}{2\pi} \int\limits_{0}^{2\pi} d \phi \left( E_1 - \bar{E_1} \right)^{2n}= \frac{(2n-1)!!}{2^n n!} \left( \tilde{A}^{2n} + \tilde{B}^{2n} \right) \nonumber \\
  + \sum\limits_{p=1}^n \sum\limits_{q=0}^{n-p} \sum\limits_{s=0}^{p-1} \frac{(-1)^{n+q+s-1}}{2^{2n-1}} \tilde{A}^{2n-2p+1} \tilde{B}^{2p-1} \left(\begin{array}{c} 2n \\ 2p-1 \end{array} \right) \left( \begin{array}{c} 2n-2p+1 \\ q \end{array} \right) \nonumber \\ 
\times \left( \begin{array}{c} 2p-1 \\ s \end{array} \right) \cos{\left[ \left( 2n - 2p -2q +1 \right)\psi_1 - \left( 2p - 2s -1 \right)\psi_2 \right]} \nonumber \\
\times \delta_{2\left( 2n - 2p -2q +1 \right), 2p-2s-1} + \sum\limits_{p=1}^{n-1} \sum\limits_{q=0}^{n-p-1} \sum\limits_{s=0}^{p-1} \frac{(-1)^{n-q-s}}{2^{2n-1}} \tilde{A}^{2n-2p} \tilde{B}^{2p} \nonumber \\ 
\times \left( \begin{array}{c} 2n \\ 2p \end{array} \right) \left( \begin{array}{c} 2(n-p) \\ q \end{array} \right) \left( \begin{array}{c} 2p \\ s \end{array} \right) \cos{\left[ 2\left( n-p-q  \right)\psi_1 - 2\left( p-s \right)\psi_2 \right]} \nonumber \\
\times \delta_{2(n-p-q), p-s} + \sum\limits_{p=1}^{n-1} \frac{1}{2^{2n}} \tilde{A}^{2n-2p} \tilde{B}^{2p} \left( \begin{array}{c} 2n \\ 2p \end{array} \right) \left( \begin{array}{c} 2(n-p) \\ n-p \end{array} \right) \left( \begin{array}{c} 2p \\ p \end{array} \right).
\end{eqnarray}
Calculation of the odd moments is performed in the same way. Namely, 
\begin{eqnarray}\label{B4}
  \overline{\left( E_1  - \bar{E_1}\right)^{2n-1}} = \frac{1}{2\pi} \int\limits_0^{2\pi} d \phi \left( E_1 - \bar{E_1}  \right)^{2n-1} \nonumber \\
= \frac{1}{2\pi} \int\limits_0^{2\pi} d \phi \left\{ \tilde{A} \sin{\left( 2 \phi + \psi_1  \right)} + \tilde{B} \sin{\left( \phi + \psi_2  \right)}  \right\}^{2n-1}.
\end{eqnarray}
Applying Newton's binomial formula and expanding the powers of the sines~\cite[1.320]{GR} we obtain
\begin{eqnarray}\label{Bodd}
  \overline{\left( E_1 - \bar{E_1} \right)^{2n-1}} = \frac{1}{2\pi} \sum\limits_{p=1}^{n-1} \tilde{A}^{2n-2p} \tilde{B}^{2p-1} \left( \begin{array}{c} 2n-1 \\ 2p-1 \end{array} \right) \int\limits_0^{2\pi} d \phi \frac{1}{2^{2(n-p)}} \nonumber \\
\times \left\{\sum\limits_{q=0}^{n-p-1} (-1)^{n-p-q} 2 \left( \begin{array}{c} 2(n-p) \\ q \end{array}  \right) \cos{\left[ 2 \left( n-p-q  \right) \left( 2 \phi + \psi_1  \right)  \right]} \right. \nonumber \\
\left.  + \left( \begin{array}{c} 2(n-p) \\ n-p \end{array} \right) \right\} \frac{1}{2^{2p-2}} \sum\limits_{s=0}^{p-1} (-1)^{p+s-1} \left( \begin{array}{c} 2p-1 \\ s \end{array} \right) \sin{\left[ \left( 2p-2s-1  \right) \left( \phi + \psi_2  \right)  \right]} \nonumber\\
+ \frac{1}{2\pi} \sum\limits_{p=1}^{n-1} \tilde{A}^{2n-2p-1} \tilde{B}^{2p} \left( \begin{array}{c} 2n-1 \\ 2p \end{array} \right) \frac{1}{2^{2(n-p)-2}}\sum\limits_{q=0}^{n-p-1} (-1)^{n-p+q-1} \left( \begin{array}{c} 2(n-p)-1 \\ q \end{array} \right) \nonumber \\
\times \int\limits_0^{2\pi} d \phi \sin{\left[ \left( 2n-2p-2q-1 \right) \left(2 \phi + \psi_1 \right)  \right]} \frac{1}{2^{2p}} \nonumber \\
\times \left\{ \sum\limits_{s=0}^{p-1} (-1)^{p-s} 2 \left( \begin{array}{c} 2p \\ s \end{array} \right) \cos{\left[ 2\left( p-s \right)\left( \phi + \psi_2 \right)  \right]} + \left( \begin{array}{c} 2p \\ p \end{array}  \right)  \right\}.
\end{eqnarray}
where we took into account that the definite integral of the odd power of the sine taken over the sine period is zero. The remaining integrations give the result:
\begin{eqnarray}\label{BoddMoments}
  \overline{\left( E_1 - \bar{E_1} \right)^{2n-1}} = \sum\limits_{p=1}^{n-1} \sum\limits_{q=0}^{n-p-1} \sum\limits_{s=0}^{p-1} \frac{(-1)^{n+s-q}}{2^{2n-2}} \tilde{A}^{2n-2p} \tilde{B}^{2p-1} \left( \begin{array}{c} 2n-1 \\ 2p-1 \end{array} \right) \nonumber \\
\times \left( \begin{array}{c} 2(n-p) \\ q \end{array} \right) \left( \begin{array}{c} 2p-1 \\ s \end{array} \right) \sin{\left[ 2\left(n-p-q \right)\psi_1 - \left( 2p-2s-1 \right)\psi_2 \right]} \nonumber \\
\times \delta_{4(n-p-q), 2p-2s-1} + \sum\limits_{p=1}^{n-1} \sum\limits_{q=0}^{n-p-1} \sum\limits_{s=0}^{p-1} \frac{(-1)^{n+q-s-1}}{2^{2n-2}} \tilde{A}^{2n-2p-1} \tilde{B}^{2p} \nonumber \\
\times \left( \begin{array}{c} 2n-1 \\ 2p \end{array} \right) \left( \begin{array}{c} 2(n-p) -1 \\ q \end{array} \right) \left( \begin{array}{c} 2p \\s \end{array} \right) \nonumber \\
\times \sin{\left[ \left( 2n -2p -2q -1 \right) \psi_1 - 2\left( p-s \right)\psi_2 \right]}  \delta_{2n-2p-2q-1, p-s}.
\end{eqnarray}
Thus we have analytically derived the exact expressions for the even and odd energy distribution moments of arbitrary order.

\end{document}